\documentclass{jaa}

\usepackage[T1]{fontenc}
\usepackage{natbib}

\usepackage{mathptmx}
\usepackage{txfonts}
\usepackage{ae,aecompl,amsmath,amsbsy}
\usepackage{epstopdf}
\usepackage[draft]{hyperref}
\usepackage{graphicx}   
\usepackage{enumitem,url}       
\usepackage{amssymb}    
\usepackage{caption}
\usepackage{subcaption}
\usepackage{mathpazo}
\usepackage{rotating}
\usepackage{comment}
\usepackage{multirow}
\usepackage{mathcomp}
\usepackage{fixltx2e}

\usepackage{graphicx}

\def\sp#1{\textsuperscript{#1}}
\def\sb#1{\textsubscript{#1}}
\def\sbsp#1#2{\rlap{\sb{#1}}\sp{#2}}
\def\m{\textminus}

\begin{document}\sloppy

\title{Coronae of an active fast rotator FR Cnc}


\author{Jeewan C. Pandey\textsuperscript{1}, Gurpreet Singh\textsuperscript{1},  Subhajeet Karmakar\textsuperscript{1}, Arti Joshi\textsuperscript{1,2}, I. S. Savanov\textsuperscript{3}, S. A. Naroenkov\textsuperscript{3}, M. A. Nalivkin\textsuperscript{3}}
\affilOne{\textsuperscript{1}Aryabhatta Research Institute of Observational Sciences (ARIES), Manora peak, Nainital, 263001, India\\}
\affilTwo{\textsuperscript{2}School of Physics and Technology, Wuhan University, Wuhan 430072, China \\ }
\affilThree{\textsuperscript{3} Institute of Astronomy, Russian Academy of Sciences (INASAN), Pyatnitskaya ul. 48, Moscow, 119017 Russia.}


\twocolumn[{

\maketitle


\msinfo{****}{****}

\begin{abstract}
We present the first detailed X-ray study and simultaneous optical observations of the active fast rotating star FR Cnc.  The X-ray spectra are found to be explained by two-temperature  plasma model with temperature of cool and hot components of 0.34 and 1.1 keV, respectively. The X-ray light curve in the 0.5 -- 2.0 keV energy band is found to be rotationally  modulated  with the degree of rotational modulation of $\sim$ 17\%. We have also found that the X-ray light curve is anti-correlated with the optical light and colour curves in the sense that maximum X-ray light corresponds to minimum optical light and cooler region on the surface of FR Cnc. The X-ray luminosity of FR Cnc is found to be almost consistent in the last 30 years with an average value of  4.85$\times$10\sp{29} erg s\sp{-1}  in the 0.5 -- 2.0 keV energy band.
\end{abstract}

\keywords{Star---Late-type --- Active---X-ray---Individual (FR Cnc).}

}]


\doinum{***********}
\artcitid{\#\#\#\#}
\volnum{000}
\year{0000}
\pgrange{1--}
\setcounter{page}{1}
\lp{1}

\section{Introduction}
 Stars with an outer convective envelope usually show magnetic field induced activities like dark spots on the surface, flares, activity cycles, and hot outer atmosphere (i.e. chromospheric and coronal emission). These activities produce a range of emission from X-ray to radio domain \citep[e. g. see the review articles by][]{2009A&ARv..17..251S,2004A&ARv..12...71G,2003SSRv..108..577F}. The activity phenomena are found  to be more in stars with faster rotation periods.
However, for the rapidly rotating active stars, the magnetic activities reach the saturation level where the X-ray emission becomes independent of the rotation period \citep[e.g.][]{2003A&A...397..147P,2011ApJ...743...48W}. The ratio of X-ray to  bolometric luminosity (L\sb{X}/L\sb{bol}) at the saturation level is generally found as $\sim$10$^{-3}$ with saturation period between $\sim$1 to $\sim$ 4 days  for different type of active stars  \citep[see][]{2003A&A...397..147P}. Various explanations have been proposed for the saturation in active stars e.g. internal dynamo itself saturates and produces no more magnetic flux with increasing rotation \citep[e.g.][]{1987ApJ...321..958V}, maximum possible coverage of active regions on the stellar surface \citep{1984A&A...133..117V}, and a centrifugal stripping of the corona \citep{1999A&A...346..883J}. Thus rapidly rotating late-type stars are important as they display an extremely enhanced level of stellar activity at saturation when compared with that of the Sun and other slow rotating stars.

FR Cnc is one single, young, and fast rotating active star, which shows its magnetic activity at saturation level. It is a K7V-type star with the rotational period of 0.826518 d and located at a distance of $35.56\pm0.08$ pc \citep{2018A&A...616A...2L}.  It was identified as a probable active star when it was found to be an optical counterpart of X-ray source  1ES 0829+15.9 (or 1RXS J083230.9+154940) in \textit{Einstein} Slew Survey and later in  \textit{ROSAT} All-Sky Survey  \citep{1992ApJS...80..257E,1999AA...349..389V}. The X-ray flux during the \textit{ROSAT} All-Sky Survey was found to be weaker than that found in the \textit{Einstein} Slew Survey, showing its variable active nature.  Using long-term optical photometry and spectroscopy, \cite{2002IBVS.5351....1P,2005AJ....130.1231P},  for the first time, derived its rotational period to be 0.8267 days. They also showed that it consists of two long-lasting groups of spots and a presence of  H$\beta$, H$\alpha$, and Ca{\sc ii} H and K  emission lines in the optical spectra.   A few flaring events were also caught in the past in the optical band. \cite{2012MNRAS.421..132G,2007IBVS.5748....1G} detected a strong flare in B, V, R, and I filters, where peak flare flux in B band was found to be $\sim 37$ times more than the quiescent flux. Later in 2010, \cite{2018Ap.....61...30K}  reported one flare with flare duration of 32.5 minutes.   \cite{2019AstL...45..602S} also  detected a flare in 2019 with a flare energy of the order of 10\sp{33} erg.  Its short rotational period along with the highly variable chromospheric features and optical light curves imply that this star should manifest strong magnetic activity including flaring activity in X-rays.

Apart from being well studied in the optical band (photometry and spectroscopy), its coronal activities and properties are not yet probed. The log(L$_X$/L$_{bol}$) of -3 of FR Cnc indicates its high level of X-ray activity at the saturation level.  Further, it would be interesting to see its long term X-ray activity as previous X-ray observations from \textit{Einstein} observatory and ROentgen SATellite \textit{(ROSAT)} are approximately 30-40 years back. Moreover, the weaker X-ray flux during  RASS observations than that of the {\textit Einstein} observatory indicates a variable X-ray emission. In addition to that, the simultaneous X-ray and optical observations will help us to understand better any correlated photospheric and coronal emission. With this aim, we have carried out  X-ray observations using the  \textit{AstroSat} and coordinated optical observations using \textit{Zvenigorod} Observatory \citep{2018AstBu..73..344S}.    Results from ground-based optical observations have been published in  \cite{2019AstL...45..602S}.

In the forthcoming sections, we present observations and data reduction in section 2,  spectral and timing analyses in section 3,  discussion on the results obtained in section 4, and conclusions in section 5.


\begin{figure}[h]
\centering
\includegraphics[width=\columnwidth]{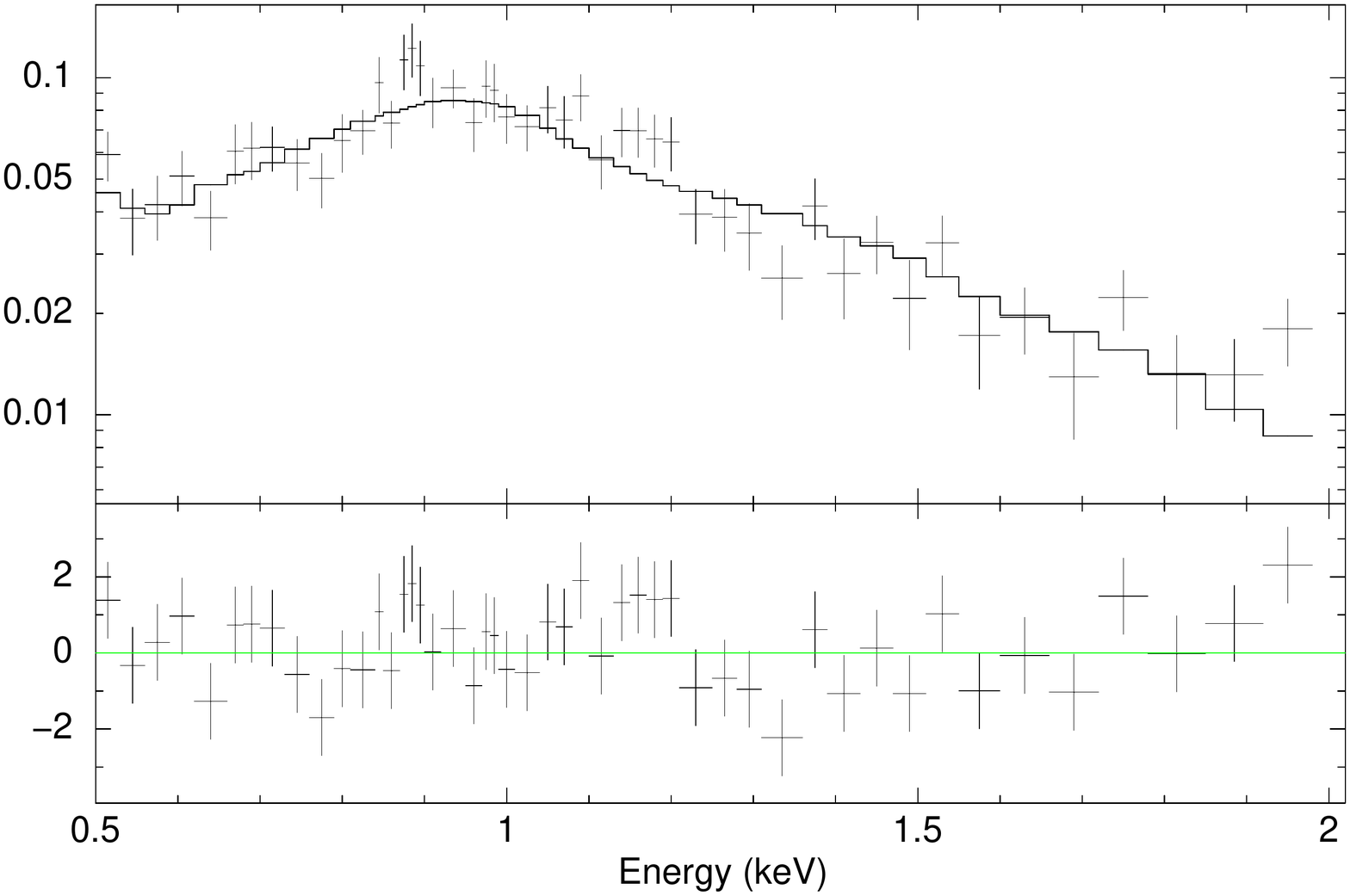}
	\caption{X-ray spectra along with the best fit 2-temperature  plasma model {\sc apec}. } 
\label{fig:spec}
\end{figure}	

\section{Observations and data reduction} \label{sec:obs}
We have observed (Observation ID. 9000002748) FR Cnc using Soft X-ray Telescope \citep[SXT;][]{2016SPIE.9905E..1ES,2017JApA...38...29S} onboard \textit{AstroSat} \citep{2014SPIE.9144E..1SS,2017JApA...38...27A}  on March 1-3, 2019. The observations were started at 18:17:08 UT on  March 1, 2019, and ended at 06:31:17 UT on   March 3, 2019. The  SXT is equipped with a CCD detector in its focal plane, which covers the energy band of  0.3 -- 8.0 keV with an effective area of $\sim$ 120 cm\sp{2} and energy resolution of $\sim$150 eV at 1.5 keV \citep{2017JApA...38...29S}. We have used Level 2 data which provides filtered events for each orbit of the \textit{AstroSat}. The Level 2 data  are collected from ISRO science data archive/\textit{AstroSat} archive\footnote{https://astrobrowse.issdc.gov.in/astro\_archive/archive/Home.jsp}. The Level 2 event files were generated by filtering  any contamination from the charged particles due to satellite's excursions through the South Atlantic Anomaly region, event grading of 0-12,  bias subtraction, and bad pixel flagging etc. Individual event files from each orbit were merged to avoid any time-overlapping events from the consecutive orbits using the {\sc sxtmergertool} provided by \textit{AstroSat} science support cell\footnote{http://astrosat-ssc.iucaa.in/?q=sxtData}.   After all the filtering the total useful exposure time was 40781 s. The X-ray spectra and light curves from the clean merged Level 2 event file were extracted using the xselect (V2.4j) package of HEASOFT\footnote{https://heasarc.gsfc.nasa.gov/docs/software/heasoft/}. In order to extract the light curve and spectra,  we have chosen the circular region of the radius 13\sp{$\prime$} centring the source for the source.  More than 90\% of source counts are found to lie within this radius. However, for the background several small source-free circular regions with a radius of 2.5\sp{$\prime$} each around the source were taken at the offset of $>$16.5\sp{$\prime$} from the centre. The response file "sxt\_pc\_mat\_g0to12.rmf" was used for the spectral analysis. 

Simultaneous optical observations of FR Cnc were planned from ARIES, Nainital and Zvenigorod Observatory of INASAN, Moscow Russia \citep{2018AstBu..73..344S}. Unfortunately, we could not observe from ARIES due to bad weather, however, we could observe in B and V bands from Zvenigorod Observatory on March 2, 3, 6, 9, 11, and 13, 2019 of which  observations carried out on March 2-3, 2019 with a total of 17.2 ks exposure were almost simultaneous with \textit{AstroSat} observations. Detailed observations, data reduction, and analysis are given in \cite{2019AstL...45..602S}.

\section{Analysis and results} \label{sec:ana}

\subsection{X-ray spectra}

\begin{table}[b]
	\caption{Best fit parameters obtained from X-ray spectral fitting. All of the errors are with the 90\% confidence interval for a single parameter.}\label{tab:spec}
	\center
	\begin{tabular}{lcc}
		\hline
		Model{$\rightarrow$}           & {\sc apec}   & {\sc apec}  \\
		Parameters{$\downarrow$}        &  (1T)  & (2T)    \\
		\hline
		kT\sb{1} (keV)               & 0.84\sbsp{\m 0.03}{+0.03}   & 0.34\sbsp{\m 0.04}{+0.06}\\
		kT\sb{2} (keV)               & ...                   & 1.1\sbsp{\m 0.1}{+0.1}\\
		EM\sb{1}(10\sp{52} cm\sp{-3})&10.7\sbsp{\m 2}{+2}       & 9.2\sbsp{\m 1}{+1}\\
		EM\sb{2}(10\sp{52} cm\sp{-3})&...                    & 4.0\sbsp{\m 1}{+1}\\
		Z (Z\sb{$\odot$})            &0.029\sbsp{ \m 0.007}{+0.007}& 0.07\sbsp{-0.02}{+0.02}\\ 
		F\sb{X}(10\sp{-12} erg s\sp{-1})&2.8$\pm$0.1             & 2.8$\pm${0.1} \\
		L\sb{X}(10\sp{30} erg s\sp{-1})&4.2$\pm$0.2             & 4.2$\pm${0.2} \\
		$\chi$\sbsp{$\nu$}{2}(dof)        & 1.3 (116)            & 1.21 (114) \\
\hline
	\end{tabular}
\end{table}

We have extracted the X-ray spectra for the energy range of 0.5 -7.0 keV.  Beyond 2.0 keV the source spectra was found to be merged with the background. Therefore, for further analysis, we have taken the X-ray spectra  from 0.5 to 2.0 keV. The X-ray spectra of FR Cnc is shown in Figure \ref{fig:spec} along with its best fit model. The X-ray spectral fitting was performed with  xspec\citep[version: 12.11.0;][]{1996ASPC..101...17A,2001ASPC..238..415D} using the $\chi^2$ statistics.  Firstly, the X-ray spectra were fitted with single temperature (1T) plasma model {\sc apec} \citep{2001ApJ...556L..91S} with solar abundance, which yielded an unacceptably high value of minimum reduced $\chi^2$ ($\chi^2_\nu$) of 3.3.  The abundances were then freed to depart from the solar values and the fit was improved significantly with  $\chi^2_\nu$ = 1.3. Further, we have fit the spectra with two temperature plasma model and found that the fit improved with  $\chi^2_\nu$ = 1.2. The best-fit parameters are given in Table \ref{tab:spec}.  For all models the solar photospheric abundances were according to \cite{1989GeCoA..53..197A}. The best-fit model yielded the plasma temperatures of 0.34 keV and 1.1 keV with abundances of 0.07Z\sb{$\odot$}, where Z\sb{$\odot$} is solar photospheric abundances. In the fitting procedure the value of hydrogen column density (N\sb{H}) was fixed to a low value of 3$\times$10\sp{20} cm\sp{2}, which is similar to the value of  N\sb{H} towards the direction of FR Cnc  \citep[see][]{1990ARA&A..28..215D}. The flux in the energy band of 0.5-2.0 keV was derived using the {\sc cflux} model. The unabsorbed flux in the energy band of 0.5-2.0 keV was found to be 4.5$\times$10\sp{29} erg s\sp{-1}.

\begin{figure} 
\includegraphics[width=\columnwidth]{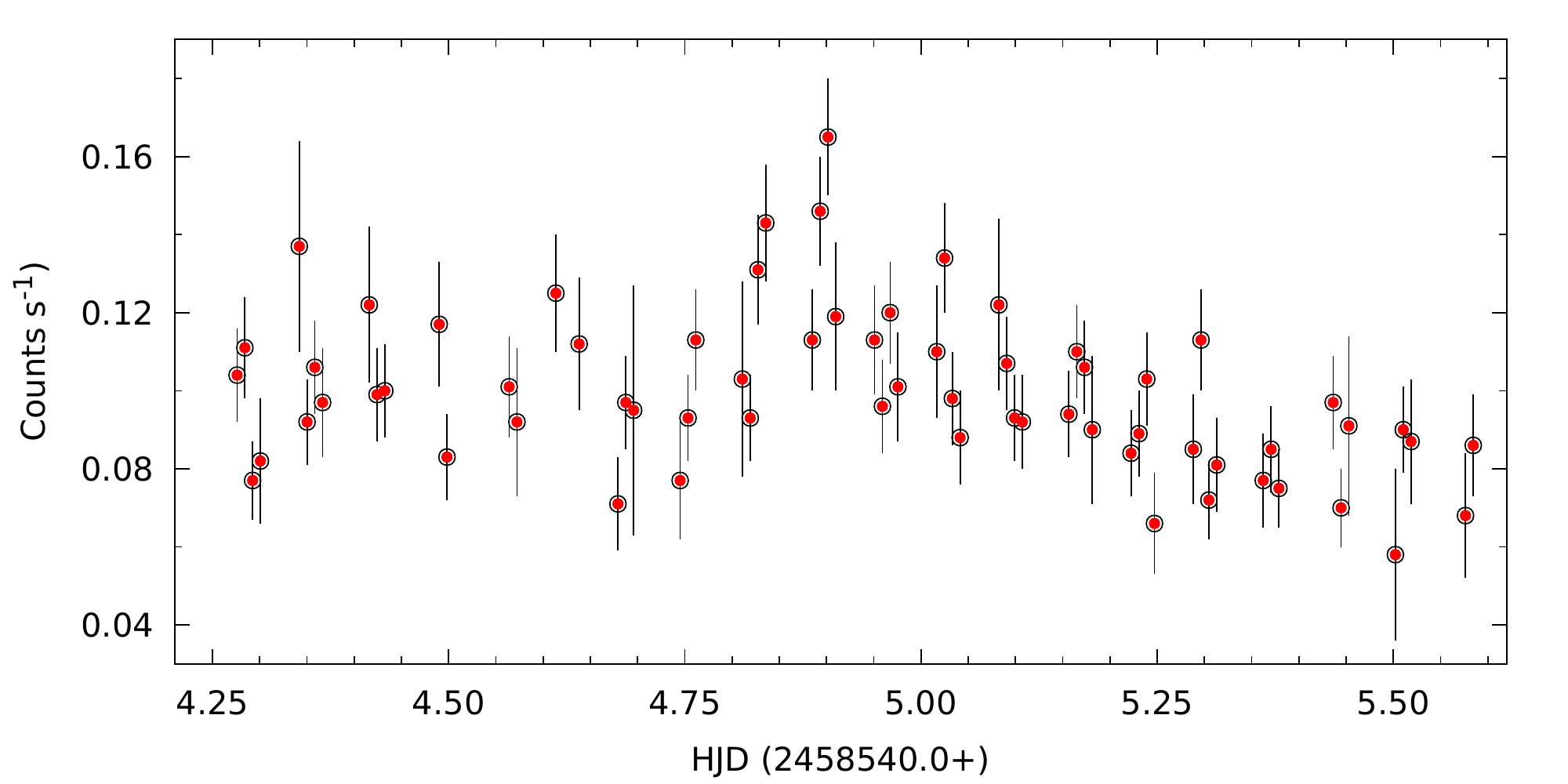}
	\caption{Background subtracted  X-ray light curve of FR Cnc in 0.5-2.0 keV energy band. }\label{fig:xlc}
\end{figure}

\subsection{X-ray light curve}
X-ray light curve was extracted in the energy band 0.5-2.0 keV. The background-subtracted X-ray light curve in the energy band 0.5-2.0 keV is shown in the Figure \ref{fig:xlc}. Binning of X-ray light curve was 711 s. The light curve appears to be variable during the observations. In order to check the variability of X-ray light curve,  we have applied $\chi^2$-test to the time series data, where  $\chi^2$ is defined as $\chi$\sp{2} = $\sum$ [\{C(t)- $\overline{C(t)}$\}\sp{2}/$\sigma$(t)\sp{2}]. Here C(t) is count rate at time t, $\overline{C(t)}$ is average count rate, and $\sigma$(t) is error on C(t).  The derived value of $\chi$\sp{2} of 168 for 79 degrees of freedom is  higher than the critical value of the  $\chi$\sp{2} of 111 for 99\% of the confidence limit. This  indicates  that the X-ray light curve is variable with confidence limit of 99\%.
\begin{figure}[h]
\includegraphics[width=\columnwidth]{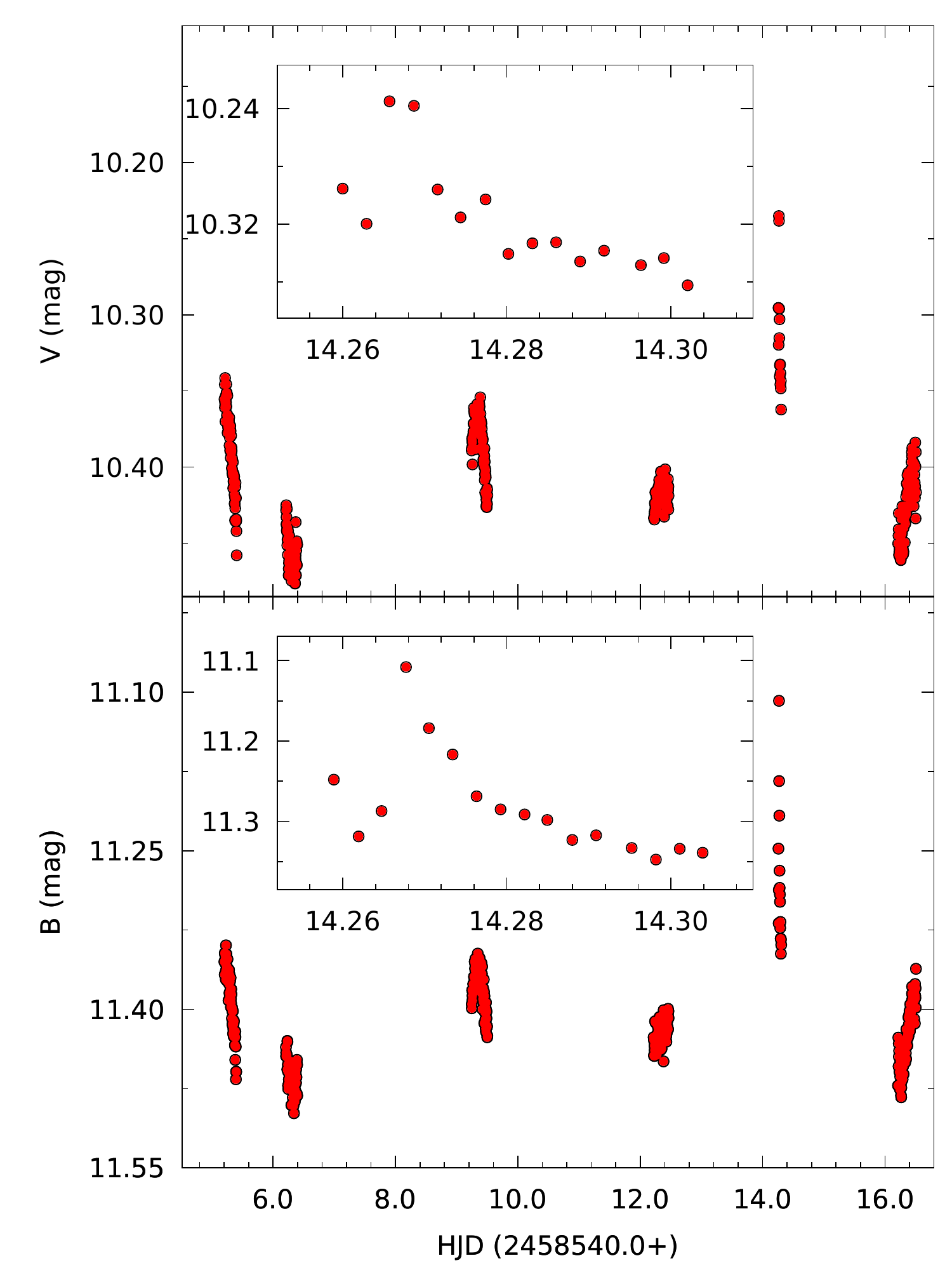}
	\caption{B and V  bands light curves of FR Cnc. Insets in each panel show the flare  light curves in the respective band  as observed on March 11, 2019.} \label{fig:olc}

\end{figure}

\begin{figure}[h]
\includegraphics[width=\columnwidth]{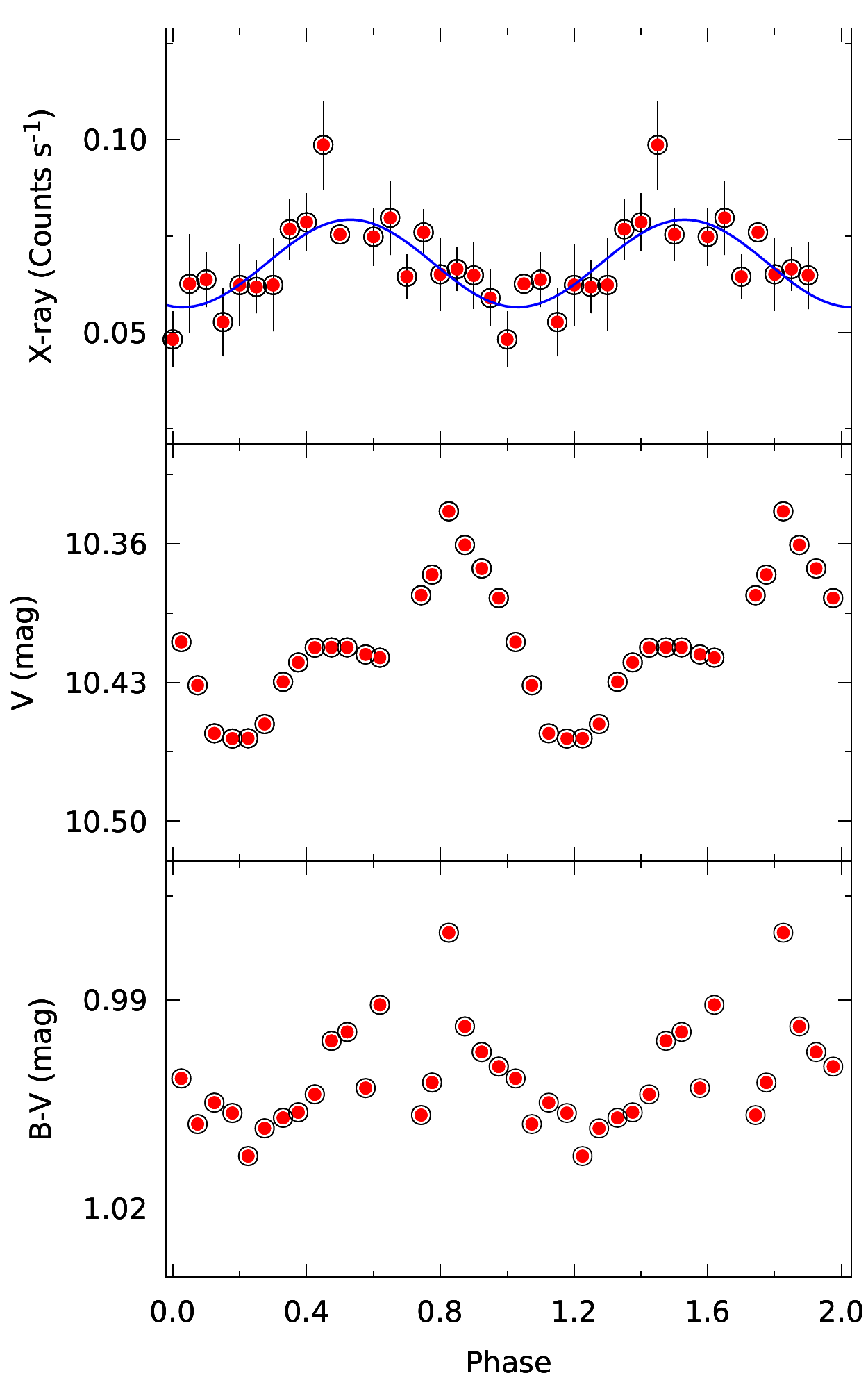}
\caption{Folded X-ray light curve of FR Cnc in X-ray and optical bands. The colour curve is shown in the bottom panel. The error in single  measurement of V is 0.008 mag. The continuous blue curve in the top panel is the best fit sinusoidal curve to the folded X-ray light curves.}\label{fig:flc}
\end{figure}

The X-ray light curve was folded with a period of 0.826518 days and at JD =  2452635.72669 as given \cite{2012MNRAS.421..132G}. The phase bin of the folded light curve was 0.05. Top panel of Figure \ref{fig:flc} shows the folded X-ray light curve in the 0.5-2.0 keV energy band.  A clear rotational modulation is present in the X-ray light curve, where peak-to-peak amplitude was found to be  $\sim$ 0.04 counts s\sp{-1}. To get the exact degree of modulation, we have fitted the folded X-ray light curve with a sinusoidal function, which yielded an amplitude of $0.011\pm0.001$ counts s\sp{-1} (see the top panel of Figure \ref{fig:flc}). This corresponds to 17$\pm$2\% modulation of the mean value of X-ray flux.

\subsection{Optical light curves}
 The optical light curves in B and V filters are shown in Figure \ref{fig:olc}. The flare observed on March 11, 2019 (HJD = 2458554.25891) is also shown in the insets. The analysis of the optical V band light curve is published in \cite{2019AstL...45..602S} where we show the surface coverage of dark spots during the observations was   $\sim$ 12\% of the total surface area of FR Cnc.   We also noticed that spots are concentrated on two active longitudes of FR Cnc. The flare was observed near phase 0.8, which is the maximum of the V-band light curve.  The flare duration was found to be $\sim$ 50 minutes in B-band whereas the total flare energy in B and V were found to be 2.2$\times$10\sp{33} and 1.4$\times$10\sp{33} erg, respectively.    

Light curve and colour curve are folded with an ephemeris as used for X-ray light curve folding. Middle and bottom panels of Figure \ref{fig:flc} shows the folded light curve in V -band and (B-V) colour curves.   We have also derived the Spearman rank correlation coefficient between X-ray and V-band fluxes, X-ray and (B-V) colour, and V and (B-V). The resultant values of the correlation coefficient with the probability of no correlation are given in Table \ref{tab:corr}. The V-band light curve was found to be strongly correlated with (B-V) colour curve in the sense that during maximum V-band magnitude the (B-V) colour is also found to be maximum. However, the X-ray flux appeared to be anti-correlated with V-band magnitude and (B-V) colour.  X-ray flux was found to be minimum near the maximum V-band flux and (B-V) colour.  
 
\begin{table}[h]
\tabularfont
\caption{Spearman rank correlation coefficients between X-ray and optical.}\label{tab:corr} 
\begin{tabular}{lcc}
\topline
	Parameters     & Correlation Coefficient & Probability\sp{*} \\
	\hline
	X-ray -- (B-V) & -0.61                   & 0.0067 \\
	X-ray -- V    & -0.54                   & 0.0020 \\
        V -- (B-V)    &  0.77                   & 0.000016 \\

	\hline
\end{tabular}
	\tablenotes{\sp{*} Probability of no correlation.}
\end{table}

\section{Discussion} \label{sec:disc}
For the first time, we present the detailed X-ray analysis and simultaneous optical observations of the active star FR Cnc. The X-ray light curve is found to be rotationally modulated. The degree of X-ray modulation was found to be 17\% in FR Cnc.  Rotational modulation in X-ray was also found in the several other active stars like  V711 Tau \citep{1988MNRAS.235..239A,2001A&A...365L.318A}, AB Dor \citep{2005ApJ...621..999H,1997A&A...320..831K} and V1147 Tau \citep{2013MNRAS.430.2154P}, where the degree of rotational modulation was found up to 30 \%. 
The maximum of the V-band light curve is found near the minimum of the X-ray light curve.  Also, we found that both V band flux and colour are anti-correlated with the X-ray flux. This indicates the high level of X-ray activity has resulted from the high magnetic regions at the surface of FR Cnc.  Also in the case of Sun, significant variability of the average X-ray flux is found to be due to the rotational modulation of the active region \citep{2004A&A...424..677O}.

 While comparing the spectral parameters of FR Cnc with other similar investigation, we found that the coronal activity of  FR Cnc lies among the other active dwarf stars.  Using the \textit{ROSAT} X-ray observations \cite{1997ApJ...478..358D} found that coronae of active dwarfs show two temperature plasma with the low-temperature component in the range of 0.13 - 0.30 keV and high-temperature component in between 1.07 to 2.81 keV. \cite{2008MNRAS.387.1627P} also found the quiescent coronae of active dwarfs consist of two temperature of 0.2-0.5 and 0.6-1.0 keV using observations from  X-ray Multi-Mirror Mission (\textit{XMM-Newton}). In the case of FR Cnc,  we found similar values of low and high temperatures. The average values of volume emission measures EM1 and EM2 for FR Cnc were found to be 9.2$\times$10\sp{52} and 4.0$\times$10\sp{52}, respectively. These values are also very close to those for similar stars \citep{1997ApJ...478..358D,2008MNRAS.387.1627P}.  The X-ray luminosity of FR Cnc is found to be similar to the average value of 4.0$\times$10\sp{29} erg s\sp{-1} for 101 active dwarf stars \citep{2005AJ....130.1231P}. Also,  FR Cnc  has a similar X-ray activity level to that of single rapidly rotating active stars \citep[e.g. LO Peg, BO Mic, AB Dor etc.;][]{2016MNRAS.459.3112K,2008ApJ...679.1509G} whereas its X-ray luminosity is less than the other fast  rotating (period $<$ 1 d) active dwarf binaries \citep[see Catalogue of Chromospherically Active Binaries;][]{2008MNRAS.389.1722E}. If we compare our results with evolved active stars, we found that FR Cnc is less active with respect to them. \cite{2012ASInC...6..239P}  showed that quiescent coronae of evolved active stars are explained by 3 temperature plasma with the median values of 0.38, 0.99, and 2.92 keV.  Also, the median value of  X-ray luminosity for evolved active stars is an order higher than that of FR Cnc and other active dwarfs. The abundance derived for FR Cnc is little lower than to that found in other active stars \citep{2001ApJ...557..747I,2008MNRAS.387.1627P,2012MNRAS.419.1219P}.

\begin{table*}[htb]
\tabularfont
\center
\caption{X-ray flux and luminosity of FR Cnc during different observations in the energy band 0.5-2.0. }\label{tab:xrl} 
\begin{tabular}{lccccc}
\topline
	Observatory    & Date of Obs.&Flux (10\sp{-12}        & L\sb{X} (10\sp{29}& Count Rate      & Reference \\
	               & (dd/mm/yyyy)&erg s\sp{-1} cm\sp{-2}) & erg s\sp{-1})     &(counts s\sp{-1}) &      \\ 
\hline
	\textit{Einstein}       &1978-1981     &10.9$\pm$4.5  & 16.4 $\pm$ 6.7 & 0.56$\pm$0.23  & \cite{1992ApJS...80..257E} \\
	\textit{ROSAT}          &09-29/10/1990 & 2.5$\pm$0.6  &  3.8 $\pm$ 0.9 & 0.29$\pm$0.07  & \cite{1999AA...349..389V}\\
	\textit{XMM-Newton}     &30/10/2006    & 4.4$\pm$0.9  &  6.7 $\pm$ 1.4 & 4.00$\pm$0.86  & \cite{2008AA...480..611S} \\
..                              &05/11/2008    & 2.9$\pm$0.6  &  4.4 $\pm$ 0.9 & 2.65$\pm$0.59  & \cite{2008AA...480..611S} \\
	\textit{AstroSat}       &01-03/03/2019 & 3.0$\pm$0.1  &  4.5 $\pm$ 0.2 & 0.112$\pm$0.001  & Present study\\
	
\hline
\end{tabular}
	\tablenotes{Count rates  from different observations  are in the respective energy band of the instrument of corresponding observatories.} 
\end{table*}

\begin{figure}
\includegraphics[width=\columnwidth]{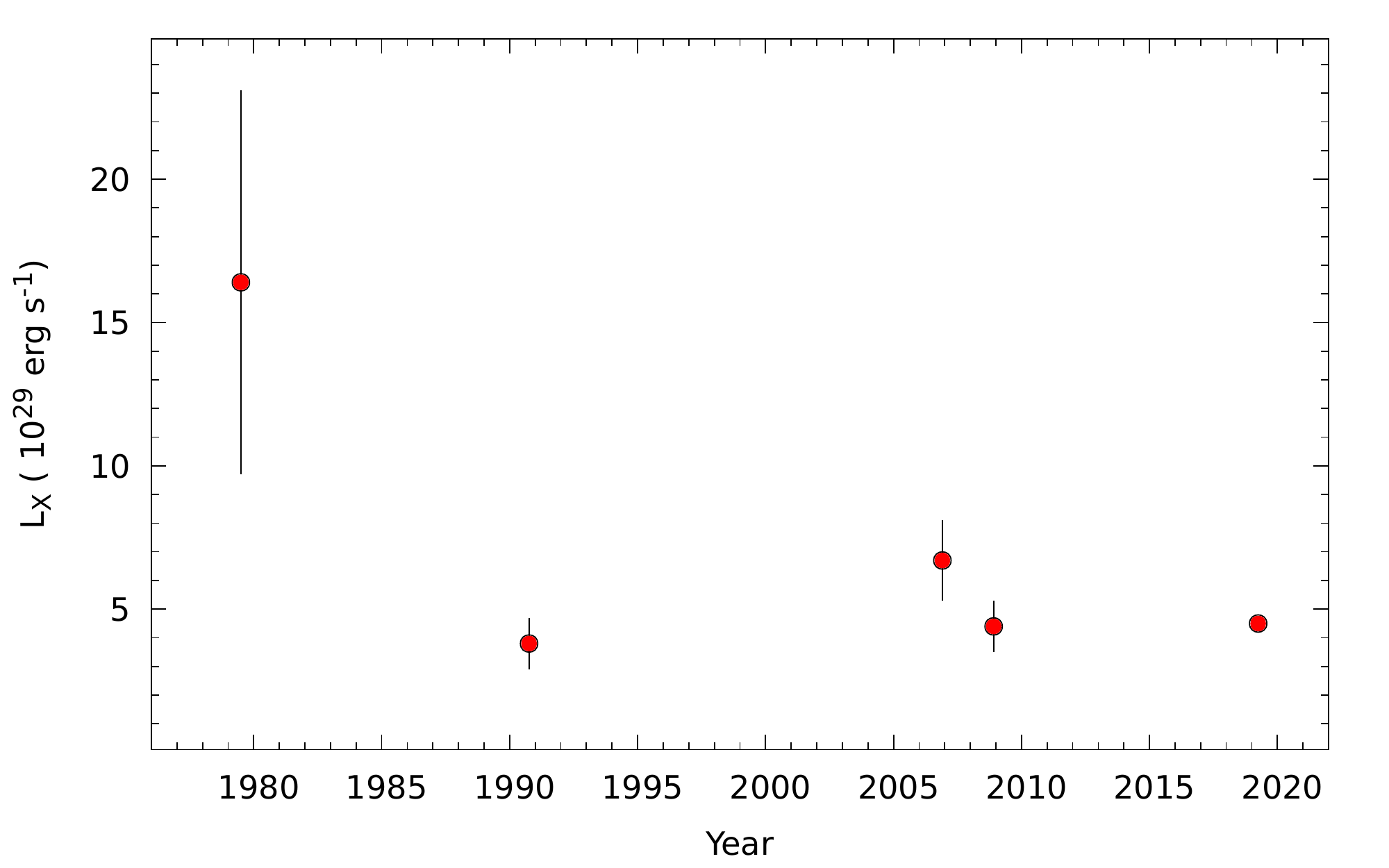}
	\caption{Long term X-ray variation of FR Cnc in 0.5-2.0 keV energy band. } \label{fig:xrl}
\end{figure}

We have also investigated the long-term behaviour of FR Cnc in the X-ray band. For this, we have taken the count rates of the past X-ray observation from  Einstein slew survey  \citep{1992ApJS...80..257E}, ROSAT all-sky survey \citep{1999AA...349..389V}, and  XMM-Newton slew survey \citep{2008AA...480..611S}.  These count rates are converted into flux using WebPIMMS\footnote{https://heasarc.gsfc.nasa.gov/cgi-bin/Tools/w3pimms/w3pimms.pl}. The flux is estimated in the 0.5-2.0 keV energy band by assuming a thermal plasma of 1.0849 keV and an abundance of 0.2 Z\sb{$\odot$}. In Table \ref{tab:xrl}, we present the count rates from these observations and corresponding  X-ray luminosity in 0.5 -2.0 keV energy band along with the flux derived from \textit{AstroSat} observations. Figure \ref{fig:xrl} shows the variation of X-ray luminosity with time. The X-ray luminosity during the \textit{AstroSat}, \textit{ROSAT} and \textit{XMM-Newton} were found to be consistent within a 1.5 $\sigma$ level indicating the almost constant level of the X-ray activity in the last 30 years. However, the X-ray luminosity of FR Cnc, $\sim$ 10 years earlier than the \textit{ROSAT} observations was 2-4 times more than the other four observations. This indicates that coronae of FR Cnc is highly active during the observations from the \textit{Einstein} observatory and became less active during later observations from the year 1990.  Due to limited data points, it is not possible to comment on any X-ray activity cycle in FR Cnc. 
There are only a few examples which show the  convincing X-ray cycles such as $\alpha$ Cen B \citep{2009ApJ...696.1931A}, HD 81809 \citep{2008A&A...490.1121F}, and  64 Cyg A \citep{2006A&A...460..261H}. All these stars have relatively low magnetic activity levels.  However, the past observational results have shown a lack (or a little-evidence) of presence of long-term  X-ray activity cycle in  highly active stars \citep[e.g.][]{1998ASPC..154..223S}. It has been suggested that the general lack of the observed cyclic activity in most active rapidly rotating stars could be due to the turbulent or distributive dynamo \citep{1996ApJ...469..828D,1999ApJ...524..988K}.  We encourage continuous  X-ray monitoring of FR Cnc to know its long term active nature.

\section{Conclusions} \label{sec:con}
The first pointed X-ray observations of the active fast rotator FR Cnc show  a variable coronal emission. It is found to be rotationally modulated with 17 \% degree of modulation around the mean flux. The X-ray variation was found to be anti-correlated with the (B-V) colour and V-band flux indicating the higher X-ray flux corresponds to the more active regions on the surface of FR Cnc. With the limited and sparse data points, the X-ray activity of FR Cnc since last 30 years appears to be almost constant. The X-ray spectra are explained by two temperatures plasma model with cooler and hotter temperatures of 0.34 and 1.1 keV, respectively. The X-ray luminosity during the \textit{AstroSat} observations is found to be 4.5$\times$10\sp{29} erg s\sp{-1} in the 0.5 -2.0 keV energy band.

\section*{Acknowledgements}
This research is based  on the results obtained from the \textit{AstroSat} mission of the Indian Space Research Organisation (ISRO), archived at the Indian Space Science Data Centre (ISSDC). This work has been performed utilizing the calibration data-bases and auxiliary analysis tools developed, maintained and distributed by AstroSat-SXT team with members from various institutions in India and abroad.  This research has been done under the Indo-Russian DST-RFBR project reference INT/RUS/RFBR/P-167, INT/RUS/RFBR/P-271 (for India) and Grant RFBR Ind a 14-02-92694, 17-52-45048 and grant N 075-15-2020-780 ( for Russia). We acknowledge the referee for reading our paper and his/her comments.
\vspace{-1em}

\bibliographystyle{apj}

\bibliography{my_ref}

\end{document}